\newcommand{\nbold}[1]{\noindent\textbf{#1}}

\newcommand{\model}{XstrAI}
\documentclass[sigconf, nonacm]{acmart}

\usepackage{tcolorbox}
\usepackage{colortbl}
\usepackage{xcolor}
\usepackage{multirow}
\usepackage{graphicx}

\makeatletter
\gdef\@acmConference{}
\gdef\@acmConference@short{}
\AtBeginDocument{%
  \gdef\@acmConference{}%
  \gdef\@acmConference@short{}%
}
\makeatother

\AtBeginDocument{%
  }

\makeatletter
\gdef\@acmConference{}
\gdef\@acmConference@short{}
\makeatother
\setcopyright{cc}
\copyrightyear{2026}
\acmYear{2026}
\acmDOI{XXXXXXX.XXXXXXX}

\acmISBN{978-1-4503-XXXX-X/2018/06}

\author{Francesco Musicco}
\affiliation{\institution{Politecnico di Bari, Italy}
  \city{}
  \country{}}
\email{francesco.musicco@poliba.it}

\author{Danilo Danese}
\affiliation{\institution{Politecnico di Bari, Italy}
  \city{}
  \country{}}
\email{danilo.danese@poliba.it}

\author{Giuseppe Fasano}
\affiliation{\institution{Politecnico di Bari, Italy}
  \city{}
  \country{}}
\email{giuseppe.fasano@poliba.it}

\author{Angela Lombardi}
\affiliation{\institution{Politecnico di Bari, Italy}
  \city{}
  \country{}}
\email{angela.lombardi@poliba.it}

\author{Alberto Carlo Maria Mancino}
\affiliation{\institution{Politecnico di Bari, Italy}
  \city{}
  \country{}}
\email{alberto.mancino@poliba.it}

\author{Tommaso {Di Noia}}
\affiliation{\institution{Politecnico di Bari, Italy}
  \city{}
  \country{}}
\email{tommaso.dinoia@poliba.it}

\begin{document}
  \settopmatter{printacmref=false}
\title{Who Are You Explaining To? \\A Multi-Agent System for Audience-Aware XAI Narratives}


\renewcommand{\shortauthors}{Musicco et al.}

\begin{abstract}
Feature-attribution methods such as SHAP provide useful evidence about individual model predictions, but their numerical outputs are rarely sufficient for audiences with different expertise, goals, and risks of misinterpretation. In medical AI, the same local explanation must reach patients, clinicians, and data scientists through markedly different forms of communication, and naive verbalization through large language models (LLMs) is prone to weak grounding, conflation of attribution with causal language, and outputs that are persuasive without being faithful to the underlying model evidence. We introduce XstrAI, an audience-aware multi-agent framework that treats local explanations as fixed evidence and structures how it is communicated to each target reader. Each prediction case is encoded as an immutable structured representation, shared identically across audiences so the underlying evidence remains fixed. Generation is factored into three specialized LLM agents responsible for audience-aware planning, linguistic realization, and validation for grounding, attribution consistency, communicative risk, and audience appropriateness, with a bounded revision loop triggered on detected inconsistencies. We evaluate XstrAI on diabetes and stroke risk prediction against 11 baselines, ranging from direct verbalization to a re-implementation of a state-of-the-art narrator. The evaluation combines an intra-narrative regime measuring fidelity to SHAP evidence with an extra-narrative regime assessing audience appropriateness through reference corpora, multi-family LLM judges, and a survey with target readers. In both evaluations, XstrAI's narratives are consistently assigned to their intended audience by independent judges, and preferred over all baselines on Clinician and Patient audiences, with competitive performance on Data Scientist, where audience-conditioned single-prompt baselines lead. We make the code publicly available at \url{https://github.com/sisinflab/XstrAI}.
\end{abstract}

\ccsdesc[100]{Computing methodologies~Natural language processing}
\ccsdesc[100]{Information systems~Collaborative and social computing systems and tools}

\keywords{XAI, Agent Based AI, LLM, SHAP, Narrative XAI}

\maketitle
\section{Introduction}
Local feature-attribution methods such as SHAP have become a standard component of post-hoc interpretability pipelines \cite{DBLP:conf/nips/LundbergL17}, providing per-instance evidence about model behavior in the form of feature contributions, ranks, and directions. As these methods are deployed in operational settings, the audience of an explanation has emerged as a determinant of its usefulness no less important than its technical fidelity: the same attribution output supports different inferences, different admissible claims, and different downstream actions depending on whether it is consumed by a domain expert, a non-expert decision subject, or a model developer \cite{DBLP:conf/chi/EhsanLMRW21, DBLP:journals/corr/abs-2110-10790, DBLP:journals/csur/NautaTPNPSSKS23, DBLP:journals/corr/abs-1810-00184}. Empirical work on human-AI collaboration further shows that explanations misaligned with the reader can induce inappropriate reliance and degrade joint performance, even when the underlying evidence is technically correct \cite{DBLP:conf/chi/BansalWZFNKRW21, DBLP:journals/inffus/LongoBCCCSGHHHJKLMPSSSS24}. Audience adaptation is therefore a condition for the safe and effective use of XAI outputs, rather than a presentational refinement.

Existing approaches address this requirement only partially. Feature-attribution methods are by design audience-agnostic, leaving the transformation of evidence into communicable content to whoever consumes the output. Direct verbalization of attribution values through large language models (LLMs), or single-prompt role conditioning, are the natural shortcut, but recent evidence indicates that such generation is prone to weak grounding \cite{DBLP:journals/csur/JiLFYSXIBMF23}, to a conflation of attribution with causal language \cite{DBLP:journals/tmlr/ZecevicWDK23}, and to outputs that are persuasive without being faithful to the underlying model evidence \cite{DBLP:journals/inffus/LongoBCCCSGHHHJKLMPSSSS24, DBLP:conf/nips/TurpinMPB23}. Narrative-XAI systems improve accessibility but largely treat generation as a monolithic step, in which evidence selection, audience adaptation, linguistic realization, and validation are entangled within a single pass and offer limited inspectable control \cite{DBLP:conf/wsse/DwiyantiWN25, DBLP:journals/dss/MartensHDVE25, DBLP:conf/bigdataconf/ZytekPABV24}.

In this work we introduce \model, a multi-agent framework for audience-aware communication of local explanation evidence. Rather than treating explanation generation as a monolithic narration task, \model~structures explanation communication as a controlled transformation pipeline grounded in a shared, immutable representation of the prediction case. Each instance is encoded as a machine-readable artifact that fixes the prediction, attribution values, feature semantics, and generation constraints identically across stakeholder pathways. Within each pathway, generation is factored into three role-specific LLM agents responsible for stakeholder-aware planning, linguistic realization, and validation against grounding, attribution consistency, communicative risk, and audience appropriateness, with a bounded revision loop triggered when inconsistencies are detected. We instantiate the framework on two clinical prediction tasks, diabetes and stroke risk estimation, where stakeholder heterogeneity is well documented and the cost of miscommunication is concrete. We evaluate \model~against 11 baseline  configurations, including a re-implementation of Explingo \cite{DBLP:conf/bigdataconf/ZytekPABV24} and a family of incremental ablations. The evaluation is organized around two complementary regimes that reflect the dual nature of the problem: an \textit{intra-narrative} regime measuring text form, audience-specific semantics and fidelity of the produced narrative to the underlying explanation evidence, and an \textit{extra-narrative} regime measuring its appropriateness for the target reader against audience-specific human reference corpora, three independent LLM judge families aggregated via a Bradley-Terry/Elo model, and a human study with representatives of each target audience. 
Our contributions are threefold:
\begin{itemize}
    \item A novel architecture for audience-aware XAI narrative generation that decouples evidence representation, stakeholder-specific planning, linguistic realization, and validation into inspectable stages.
    \item A dual evaluation protocol that disentangles intra-narrative fidelity to the underlying explanation evidence from extra-narrative appropriateness for the target audience, with the latter assessed against audience-specific human reference corpora, multi-family LLM judges, and a human study.
    \item An empirical study on diabetes and stroke prediction against  11 baseline configurations, with rank-based statistical analysis of the audience-specific outcomes.
\end{itemize}

\section{Related Work}


\subsection{XAI and Stakeholder Needs}

Local explanation methods occupy a central role in the XAI literature by providing instance-level evidence for individual model predictions. Among them, feature-attribution methods such as SHAP~\cite{DBLP:conf/nips/LundbergL17} estimate the contribution of input features to a prediction, enabling users to inspect and interpret the factors influencing the model output at the level of individual instances. These methods are widely adopted for model debugging, feature attribution analysis, and post-hoc inspection of black-box systems.

The increasing use of local explanations has progressively exposed a distinction between explanation evidence and explanation communication. Feature attribution scores, ranked feature lists, and visualization-based outputs provide technically informative representations of model behavior, but their interpretation depends on the reader's expertise, objectives, and contextual knowledge~\cite{DBLP:journals/inffus/LongoBCCCSGHHHJKLMPSSSS24, DBLP:journals/ijmms/KimKKSK24}. Prior work in human-centered XAI emphasizes that explanation quality cannot be reduced to technical properties such as fidelity or stability, since explanations operate as communicative artifacts shaped by human goals and decision-making contexts \cite{DBLP:journals/corr/abs-2110-10790, DBLP:journals/ai/Miller19}, and empirical studies on human-AI collaboration further show that explanations misaligned with the reader can induce inappropriate reliance even when the underlying evidence is technically correct \cite{DBLP:conf/chi/BansalWZFNKRW21, DBLP:journals/inffus/LongoBCCCSGHHHJKLMPSSSS24}.
This concern is particularly acute in multi-stakeholder environments, where the same predictive evidence must support different interpretive goals: clinicians typically require technically grounded inspection of feature contributions, patients require accessible and non-alarming communication, and data scientists focus on attribution behavior and model dynamics \cite{DBLP:conf/chi/EhsanLMRW21, DBLP:journals/csur/NautaTPNPSSKS23, lombardi2023human, DBLP:journals/corr/abs-1810-00184}.
Existing local explanation methods provide valuable evidence for understanding individual predictions but offer limited support for controlled stakeholder-specific communication.

\subsection{Natural Language and Narrative XAI}

Recent work has explored natural language as a means to make explanation evidence more accessible and actionable with LLM-based approaches transforming model outputs and attribution artifacts into textual explanations~\cite{bilal2025llms}. Systems such as Explingo~\cite{DBLP:conf/bigdataconf/ZytekPABV24} focus on prediction-level explanation generation, narrative-driven XAI~\cite{DBLP:journals/dss/MartensHDVE25} frames explanations as structured stories, and ContextualSHAP~\cite{DBLP:conf/wsse/DwiyantiWN25} enriches SHAP-based explanations through contextual language generation. More recent work has further emphasized personalization and trustworthiness in natural language explanations, reflecting the growing interest in explanation generation as a communication-oriented layer of XAI \cite{DBLP:journals/corr/abs-2603-06485}. This direction marks an important shift from raw numerical or visual artifacts toward more interpretable textual forms, supporting contextualization, selective emphasis, and coherent presentation of local evidence.

Fluent explanation generation, however, introduces specific risks. Textual explanations may appear coherent while weakening the connection with the underlying evidence, particularly when produced directly from attribution values without explicit validation: direct generation by LLMs can introduce hallucinations and unsupported claims \cite{DBLP:journals/csur/JiLFYSXIBMF23}, and language models are known to produce plausible explanations that misrepresent the actual factors influencing a decision \cite{DBLP:conf/nips/TurpinMPB23}, with the additional risk of conflating attribution with causal inference \cite{DBLP:journals/tmlr/ZecevicWDK23}. These observations are consistent with the broader distinction between explanations that are persuasive to users and explanations that remain substantively aligned with the model evidence \cite{DBLP:journals/inffus/LongoBCCCSGHHHJKLMPSSSS24}, and reinforce the view that explanation quality requires multiple dimensions beyond fluency \cite{DBLP:journals/csur/NautaTPNPSSKS23}. 

Existing XAI narratives approaches primarily improve the accessibility of explanation evidence while leaving open the problem of controlled generation. In particular, current systems provide limited explicit separation between evidence selection, stakeholder-specific framing, linguistic realization, and validation, motivating the need for explanation generation pipelines in which narrative production is treated as a structured and verifiable process rather than as a direct transformation from explanation values to text.

\subsection{Multi-Agent and Role-Based LLM Systems}

Recent advances in LLMs have stimulated growing interest in multi-agent and role-based interaction frameworks for complex reasoning tasks, in which language models are organized into specialized roles that interact through collaboration, critique, or coordinated task decomposition \cite{DBLP:conf/nips/LiHIKG23, DBLP:conf/uist/ParkOCMLB23}. Parallel research on role-playing and persona prompting has shown that agent behavior can be guided through explicit role assignments and evaluated in terms of behavioral consistency and task-oriented effects \cite{DBLP:conf/emnlp/AraujoRHR25, DBLP:conf/acl/WangXHYXGTFL0CL24}, while multi-agent debate frameworks demonstrate that structured interaction between language models can improve factual consistency and reduce unsupported claims by exposing disagreements and refining intermediate outputs \cite{DBLP:conf/icml/Du00TM24}. Within domain-specific reasoning, MedAgents~\cite{DBLP:conf/acl/TangZ0L0ZCG24} shows that collaborative interaction between multiple language-model agents can improve medical reasoning through distributed analysis and iterative discussion, a setting that resonates with the clinical instantiation considered in this work.

\begin{figure*}[t]
\centering
\includegraphics[width=\textwidth]{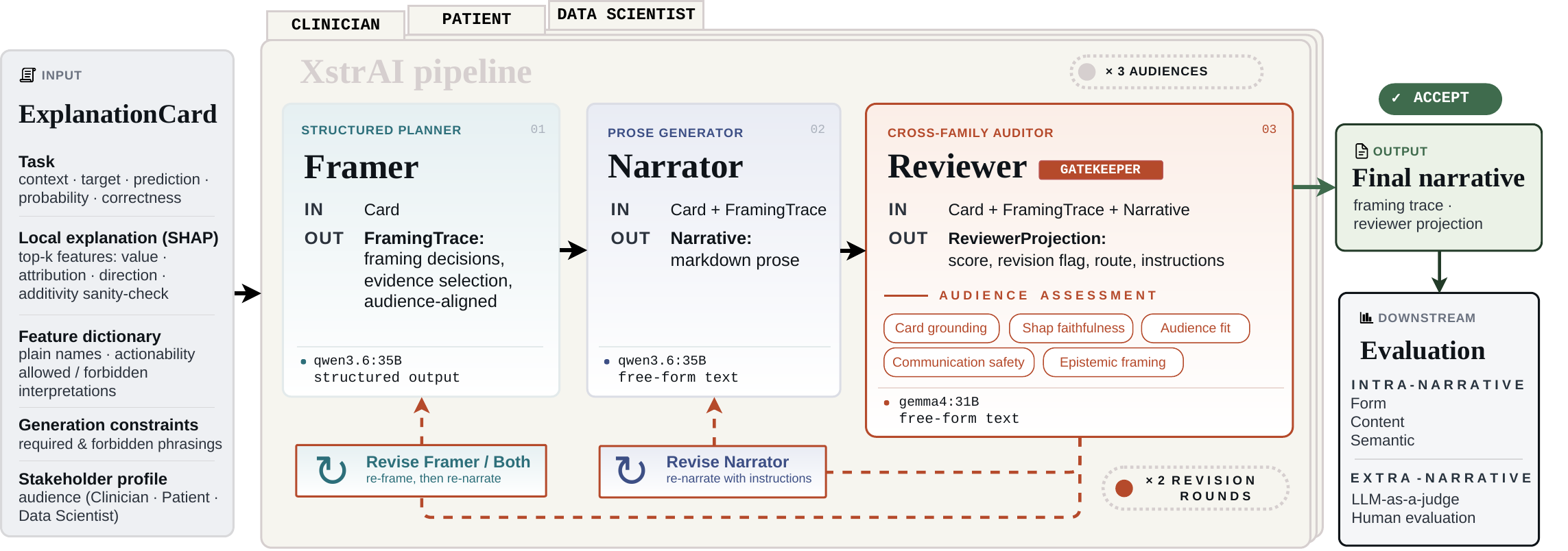}
\caption{XstrAI architecture. Each case is represented as a structured \texttt{ExplanationCard} and processed through three stakeholder-specific pathways, each comprising three agents: the \emph{Framer} produces a narrative plan, the \emph{Narrator} generates the explanation, and the \emph{Reviewer} evaluates the output across five audience-aligned dimensions. Rejected outputs trigger a bounded revision loop routed back to the Framer or Narrator. Finally, the generated narratives undergo intra- and extra-narrative evaluation.}

\label{fig:Architecture}    
\end{figure*}

\label{sec:XstrAI}

Current multi-agent systems focus on reasoning performance~\cite{DBLP:conf/icml/Du00TM24}, collaborative problem solving~\cite{DBLP:conf/iclr/HongZCZCWZWYLZR24}, behavior simulation~\cite{DBLP:conf/uist/ParkOCMLB23}, or persona fidelity~\cite{DBLP:conf/acl/WangPQLZWGGN00024}. The communication of fixed XAI evidence introduces a different requirement structure, in which the central objective is the controlled transformation of explanation evidence into stakeholder-specific narratives while preserving grounding and attribution consistency. Existing approaches provide useful foundations for role specialization and iterative refinement, while offering limited support for explanation-specific validation and explicit control over how explanatory evidence is selected, adapted, and verified throughout the generation process. 
Our work bridges these research strands by treating local explanations as fixed evidence to be communicated, and by integrating narrative XAI, stakeholder adaptation, and multi-agent control within a single explanation-communication pipeline instantiated on SHAP attributions.

\section{\model~Framework}

Figure~\ref{fig:Architecture}  illustrates \textbf{\model}, the proposed framework for transforming local feature-attribution explanations into narratives tailored to different target audiences, or stakeholders, that we use interchangeably in the rest of the paper. Operating at the explanation-communication layer, the framework keeps the predictive output and attribution values fixed, while controlling how the same evidence is selected, organized, verbalized, and validated for each target reader.

Given the clinical nature of the considered scenarios, \model~is instantiated around three stakeholder profiles: clinician, patient, and data scientist. These profiles capture different explanatory needs, requiring the same local explanation to be transformed into audience-specific narratives from a shared evidential basis.

The architecture is built around an \texttt{ExplanationCard}, which provides the structured case-level evidence and serves as source of truth for the pipeline. From this representation, three agentic components operate within each stakeholder-specific pathway: the \emph{Framer} plans the explanation, the \emph{Narrator} generates the narrative, and the \emph{Reviewer} validates the output through bounded revision cycles when needed. Intermediate and final artifacts are retained to support traceability and inspection.
The modular design is motivated by recent multi-agent LLM research highlighting the benefits of role-based task decomposition and explicit critique mechanisms~\cite{DBLP:conf/iclr/HongZCZCWZWYLZR24,DBLP:conf/icml/Du00TM24}.

\subsection{Input Representation and Control Layer}
\label{subsec:input-control-layer}

Each prediction case is represented through an \texttt{ExplanationCard}, a machine-readable source of truth for the generation process. Inspired by the Model Cards paradigm~\cite{DBLP:conf/fat/MitchellWZBVHSR19}, which standardises documentation at the \emph{model} level, the \texttt{ExplanationCard} transposes the same structured-artifact idea to the \emph{instance} level: every prediction case is paired with a self-contained, machine-readable specification of the evidence to be communicated. This specification includes case information, model output, SHAP attributions, feature metadata, generation constraints, and the stakeholder profile.

The \texttt{ExplanationCard} acts as a structured data contract between the predictive layer and the narrative generation pipeline: all stakeholder pathways receive the same validated representation, so differences across narratives depend on communicative choices rather than on changes in the underlying evidence. A deterministic controller orchestrates each pathway by passing the card to the Framer, forwarding the resulting \texttt{FramingTrace} to the Narrator, and sending the candidate narrative to the Reviewer, which either accepts the output or triggers a bounded revision step.

\subsection{Framer Agent: Stakeholder-Aware Planning}
\label{subsec:framer-agent}

The Framer is the first role-specific LLM module executed within each stakeholder pathway. It receives the shared \texttt{ExplanationCard} and transforms its evidence into a structured narrative plan, named \texttt{FramingTrace}. This artifact defines how the local explanation should be communicated before natural-language generation, including the organization of the narrative, the feature presentation order, the intended tone, the required cautions, and the operational instructions for the downstream Narrator.

Stakeholder adaptation is mainly introduced at this planning stage. For clinicians, the Framer can preserve a more medically detailed explanation; for patients, it prioritizes readability, uncertainty framing, and non-alarming formulations; for data scientists, it can retain more explicit references to attribution direction, feature ranking, relative magnitude, and model behavior. The \texttt{FramingTrace} remains an internal planning artifact rather than a user-facing explanation, thereby separating explanation planning from narrative realization.

\subsection{Narrator Agent: Controlled Narrative Generation}
\label{subsec:narrator-agent}

The Narrator is the role-specific LLM module responsible for generating the explanatory text. It receives the original \texttt{ExplanationCard}, providing the factual and attributional evidence, alongside the \texttt{FramingTrace}, which defines the stakeholder-oriented narrative plan. The output is a candidate narrative grounded in the card and constrained by the Framer's specified communicative strategy.

The generation process is controlled by the interaction between these two artifacts. The \texttt{ExplanationCard} constrains \textit{what} can be said, including the prediction, probability, selected features, observed values, attribution values, and contribution directions. The \texttt{FramingTrace} constrains \textit{how} this evidence should be communicated, including order, level of detail, tone, terminology, and cautions. This division keeps the narrative anchored to the original evidence while adapting its presentation to the target stakeholder.

The Narrator must preserve attribution meaning and respect the feature-level interpretations and generation constraints encoded in the \texttt{ExplanationCard}, including restrictions on causal, diagnostic, prognostic, and treatment-oriented claims. Its output remains provisional until it is evaluated by the Reviewer, preserving a separation between narrative generation and validation.

\subsection{Reviewer Agent and Traceability}
\label{subsec:reviewer-traceability}

The Reviewer acts as the internal quality-control module for each stakeholder-specific pathway, receiving the \texttt{ExplanationCard}, the \texttt{FramingTrace}, and the candidate narrative before final acceptance. Its assessment covers five shared dimensions across stakeholder-specific reviewers (top box in Figure~\ref{fig:reviewer_example}): card grounding, SHAP faithfulness, epistemic framing, communication safety, and audience fit. Together, these checks are designed to flag potential deviations from the input evidence, attribution meaning, epistemic boundaries, communication-safety constraints, and stakeholder-appropriate register.
For each candidate narrative, the Reviewer produces a structured assessment (bottom box in Figure~\ref{fig:reviewer_example}) that includes an acceptance judgment, a score, a rationale, optional revision instructions, and, when needed, routing recommendations for refinement. Accepted narratives are returned as final outputs, while rejected ones trigger a bounded refinement loop involving the Framer, the Narrator, or both, depending on the identified issues. Updated artifacts are subsequently re-evaluated by the Reviewer, and the process iterates until acceptance or until the maximum number of revision rounds is reached. In the current implementation, each pathway can undergo at most two refinement iterations, yielding up to three generated narratives for the same case. This bound is consistent with empirical observations on multi-agent LLM systems, where consensus across iterations becomes substantially stable within a few rounds~\cite{DBLP:conf/icml/Du00TM24}. 

Distinct from the external evaluators used in the experimental assessment, the Reviewer serves as an internal quality-control component during inference. To support traceability, the framework logs the intermediate and final artifacts generated during execution, including inputs, narratives, review outputs, revision decisions, and final accepted explanations. This trace makes it possible to reconstruct the generation process, inspect failure modes, compare stakeholder pathways, and audit the system beyond the final text.

\definecolor{xstraiSienna}{RGB}{139,62,38}
\definecolor{xstraiSiennaSoft}{RGB}{253,244,239}

\begin{figure}[t]
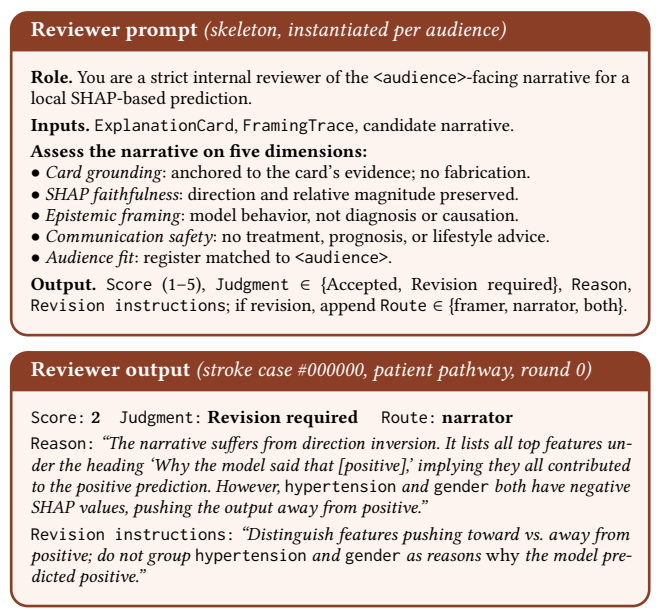

\centering

\begin{tcolorbox}[colback=xstraiSiennaSoft, colframe=xstraiSienna, arc=2mm,
    boxrule=1pt, left=4pt, right=4pt, top=4pt, bottom=4pt,
    coltitle=white,
    title=\textbf{Reviewer prompt} \emph{(skeleton, instantiated per audience)},
    fonttitle=\small, width=\columnwidth]
\footnotesize
\textbf{Role.}~You are a strict internal reviewer of the \texttt{<audience>}-facing narrative for a local SHAP-based prediction.\\[2pt]
\textbf{Inputs.}~\texttt{ExplanationCard}, \texttt{FramingTrace}, candidate narrative.\\[2pt]
\textbf{Assess the narrative on five dimensions:}\\
$\bullet$ \emph{Card grounding}: anchored to the card's evidence; no fabrication.\\
$\bullet$ \emph{SHAP faithfulness}: direction and relative magnitude preserved.\\
$\bullet$ \emph{Epistemic framing}: model behavior, not diagnosis or causation.\\
$\bullet$ \emph{Communication safety}: no treatment, prognosis, or lifestyle advice.\\
$\bullet$ \emph{Audience fit}: register matched to \texttt{<audience>}.\\[2pt]
\textbf{Output.}~\texttt{Score} (1–5), \texttt{Judgment} $\in$ \{Accepted, Revision required\}, \texttt{Reason}, \texttt{Revision instructions}; if revision, append \texttt{Route} $\in$ \{framer, narrator, both\}.
\end{tcolorbox}

\begin{tcolorbox}[colback=xstraiSiennaSoft, colframe=xstraiSienna, arc=2mm,
    boxrule=1pt, left=4pt, right=4pt, top=4pt, bottom=4pt,
    coltitle=white,
    title=\textbf{Reviewer output} \emph{(stroke case \#000000, patient pathway, round 0)},
    fonttitle=\small, width=\columnwidth]
\footnotesize
\texttt{Score:} \textbf{2}\quad\texttt{Judgment:} \textbf{Revision required} \quad\texttt{Route:} \textbf{narrator}\\[2pt]
\texttt{Reason:} \textit{``The narrative suffers from direction inversion. It lists all top features under the heading `Why the model said that [positive],' implying they all contributed to the positive prediction. However,} \texttt{hypertension} \textit{and} \texttt{gender} \textit{both have negative SHAP values, pushing the output away from positive.''}\\[2pt]
\texttt{Revision instructions:} \textit{``Distinguish features pushing toward vs.\ away from positive; do not group} \texttt{hypertension} \textit{and} \texttt{gender} \textit{as reasons \emph{why} the model predicted positive.''}

\end{tcolorbox}

\caption{The Reviewer in action. \textit{Top:} skeleton of the audience-conditioned Reviewer prompt; the same five dimensions are instantiated for Clinician, Patient, and Data Scientist. \textit{Bottom:} a real Reviewer output that catches a SHAP direction inversion in a patient narrative and routes the revision back to the Narrator, preserving the existing \texttt{FramingTrace}.}
\label{fig:reviewer_example}
\end{figure}

\section{Experimental Setting}
We conducted extensive experiments to evaluate our proposed approach. This section details the implementation of XstrAI, the selected baselines, and the evaluation protocol. All experiments were implemented in Python and executed on an Ubuntu 22.04 workstation equipped with an NVIDIA H100 GPU. The framework is model- and backend-agnostic: agent roles are defined at the prompt and schema level and can be served by any LLM endpoint exposing chat completions, including remote APIs. To support reproducibility and external inspection, the repository includes the full framework implementation, stakeholder-specific prompts, reviewer configurations, evaluation pipelines, and experimental settings: ~\url{https://anonymous.4open.science/r/XstrAI-5528}.

\subsection{Dataset, Predictive Models and Explanations}

Our experiments involved two publicly available classification datasets: the Diabetes Dataset\footnote{\url{https://www.kaggle.com/datasets/mathchi/diabetes-data-set}} and the Stroke Prediction Dataset\footnote{\url{https://www.kaggle.com/datasets/fedesoriano/stroke-prediction-dataset}}. These datasets were selected because they offer interpretable tabular features, and a manageable feature space for assessing how local explanations can be transformed into stakeholder-specific narratives, consistently with recent work on XAI narrative generation~\cite{he2026agentic}.

The Diabetes Dataset comprises 768 records from the National Institute of Diabetes and Digestive and Kidney Diseases, with eight features and a binary diagnosis. The Stroke Prediction Dataset contains 5110 records with 10 features predicting stroke risk; preprocessing included one-hot encoding for categorical features (\textit{gender}, \textit{ever\_married}, \textit{Residence\_type}), median imputation of missing BMI values, and discarding the \textit{smoking\_status}  feature due to excessive missing data. For both datasets we addressed class imbalance by randomly undersampling the majority class (yielding 536 and 498 samples respectively), applied an $80\%/20\%$ train-test split, and normalized all features to the $[0,1]$ range.


For the classification task, we trained a Random Forest (RF) model on each dataset with default hyperparameters, achieving an accuracy of 0.68 (diabetes) and 0.77 (stroke). To interpret RF predictions, we computed local SHAP attributions \cite{DBLP:conf/nips/LundbergL17},
selected for their theoretical guarantees
\cite{flora2022comparing, DBLP:conf/nips/LundbergL17} and its extensive validation in the medical domain \cite{sun2025explainable}. 
We used the \textit{TreeExplainer} algorithm from the Python SHAP library \footnote{\url{https://shap.readthedocs.io/en/latest/}}, which provides efficient computation for tree-based models. 


\subsection{Baselines}
\label{sec:exp:baselines}


We compare \model~against five baselines: Explingo~\cite{DBLP:conf/bigdataconf/ZytekPABV24}, the most recent loop-based narrator for tabular SHAP attributions, and four controlled variants B1--B4 of \model~that incrementally introduce the components of our pipeline, given the limited availability of directly comparable systems for audience-aware narrative generation from fixed local attribution evidence.

\nbold{Explingo.} Explingo~\cite{DBLP:conf/bigdataconf/ZytekPABV24} introduces a dual-agent framework consisting of a narrator and a reviewer, which the original authors refer to as a grader. First, the narrator converts an explanation tuple into a short text using three labeled few-shot exemplars. Subsequently, the reviewer evaluates this generated output across four specific rubrics: \texttt{accuracy}, \texttt{completeness}, \texttt{fluency}, and \texttt{conciseness}. If the resulting average score falls below a predefined threshold, the reviewer automatically triggers a rewrite process.
To enable a fair evaluation, we re-implement it under the same setup adopted for \model, pairing a \texttt{qwen} narrator with a \texttt{gemma} reviewer, instead of using OpenAI APIs. Prompt assembly, rubric wording and the deterministic conciseness formula follow the original release. The framework is audience-agnostic at every level: it returns a single narrative per instance, the narrator-reviewer exchange is limited to a scalar score and short free-text feedback, and the rubrics do not encode any reader-specific criterion. These properties motivate the controlled baselines that follow.

\nbold{Incremental baselines B1-B4.}
We implement a family of incremental baselines that ablate the individual contributions of \model, with consecutive baselines differing by a single design decision so as to isolate its effect. \textbf{B1} is the minimal configuration: a single language model receives the explanation card and produces one generic narrative, with neither audience conditioning nor any revision step. \textbf{B2} introduces the simplest form of audience awareness, a role description embedded in the prompt, while preserving the single-shot character of B1. \textbf{B3} adds a revision loop on top of B1: a narrator generates, a separate reviewer emits a $1$-$5$ score together with an \texttt{accept}/\texttt{revise} verdict and a short comment, and the narrator rewrites until acceptance or until a fixed budget is exhausted. \textbf{B4} extends B3 with a rule-based verifier that examines machine-readable properties of the candidate narrative, section structure, presence of the numeric SHAP values in the prose, absence of list-like or JSON-like formatting, and a rewrite is requested unless both the reviewer and the verifier accept the output. 

\nbold{Implementation details.} Every system consumes the same predictive model output and the same SHAP attribution, i.e. the same \texttt{ExplanationCard}, and every language model is instantiated from either \texttt{Qwen3.6:35B}~\cite{qwen3} or \texttt{Gemma4:31B}~\cite{gemma4modelcard}. For single-agent baselines (B1 and B2), we instantiate two versions of each system, one for each type of backbone, so that the comparison does not depend on a single language model. For configurations involving a Reviewer Agent (\model, Explingo, B3, and B4), we use \texttt{Qwen3.6:35B} as the generator and \texttt{Gemma4:31B} as the reviewer, so as to avoid potential family bias during the review process~\cite{DBLP:conf/nips/PanicksseryBF24,DBLP:conf/nips/ZhengC00WZL0LXZ23,DBLP:conf/acl/WangLCCZLCKLLS24}. 
When accounting for backbone and audience variants, the five baseline families expand into 11 distinct configurations, which together with the three audience-specific \model~variants constitute the 14-system pool used in the evaluation.

We denote systems as \texttt{system-backbone:audience}, where backbones are \texttt{qwen} or \texttt{gemma} and audience codes are \emph{C} (Clinician), \emph{P} (Patient), or \emph{DS} (Data Scientist). Audience-agnostic or single-backbone systems omit the respective segments. For instance, B1-qwen denotes B1 instantiated on qwen, while B2-gemma:P denotes B2 instantiated on gemma for the Patient audience.


\subsection{Evaluation framework}
Audience-aware XAI narratives are not yet supported by a dedicated evaluation protocol covering both their observable textual properties and how they are received by external evaluators. We therefore designed a dedicated framework articulated along two regimes: an Intra-narrative module, computing observable properties of the narratives themselves, and an Extra-narrative module, eliciting judgments from external evaluators.

\subsubsection{Intra-narrative Evaluation Module}
This module comprises a diverse set of metrics to evaluate the narratives along three distinct axes: form, content, and semantic.

\nbold{Form.}
To quantitatively assess the stylistic properties of the generated narratives, we performed a form-oriented analysis focusing on readability, lexical richness, verbosity, and repetition patterns. The objective of this analysis is not to evaluate factual correctness, but rather to characterize the structural and stylistic properties of the generated narratives, analyzing how different systems organize explanations and whether these characteristics change when narratives are adapted to specific stakeholders.

Specifically, we compute four complementary metrics. 
The average number of generated words per narrative (\textit{\#Words}) is used to estimate verbosity and explanation granularity, indicating how concise or detailed the generated explanations are.
Readability is evaluated through the \textit{Flesch--Kincaid Grade Level (FKGL)}~\cite{flesch1948new}, which estimates the educational level required to understand a text, with lower values indicating simpler language and higher values greater linguistic complexity.
To evaluate lexical richness, we compute the \textit{Moving-Average Type--Token Ratio (MATTR)}~\cite{DBLP:journals/jql/CovingtonM10}, a lexical diversity metric robust to text-length variation, where higher values indicate richer language use.
Finally, repetitive generation patterns are quantified through the \textit{Repeated Trigram Ratio (Rep-3)}~\cite{DBLP:conf/aaai/FuLSS21}, defined as the proportion of repeated three-word sequences appearing within a narrative. Higher values indicate stronger reliance on repetitive or templatic phrasing, while lower values suggest more diverse and fluent text generation.

All metrics are computed independently for each generated narrative and subsequently averaged across both datasets. Since some systems explicitly support stakeholder-oriented generation while others produce only generic explanations, the analysis additionally investigates whether stylistic properties vary when narratives are adapted to clinician, data scientist, or patient audiences.

\nbold{Content.} 
To evaluate the content faithfulness of the generated narratives, we compare the information extracted from each narrative against the corresponding \texttt{ExplanationCard}, which provides the reference feature values, SHAP contributions, and feature ordering.

We compute three agreement metrics introduced by~\citet{DBLP:journals/tmlr/KrishnaHG0JL24}. \textit{Value Agreement (VA)} measures whether the feature values mentioned in the narrative match those in the \texttt{ExplanationCard}. \textit{Sign Agreement (SA)} measures whether the direction of each feature contribution is preserved, i.e., whether a feature is described as raising or lowering the positive class consistently with the sign of its SHAP value. \textit{Rank Agreement (RA)} measures whether the ordering of features in the narrative is consistent with the original feature ordering in the \texttt{ExplanationCard}.

To compute these metrics, we use two complementary extraction strategies. The first is a deterministic rule-based extractor that identifies feature mentions through canonical names, normalized variants, and predefined aliases. For each detected feature, the extractor analyzes a local text window to identify feature values, SHAP values, contribution direction, and rank indicators. Directions are inferred either from explicit SHAP values or from lexical markers such as \textit{raises}, \textit{lowers}, \textit{positive contribution}, and \textit{negative contribution}. Feature and SHAP values are extracted through numeric regular expressions and matched against the expected values in the \texttt{ExplanationCard}. Ranks are inferred from explicit ordinal markers such as \textit{first}, \textit{second}, or \textit{most important}; otherwise, ranks are assigned according to feature appearance order.

The second strategy uses an LLM-based extractor based on \texttt{GPT-5.4-mini}, which infers feature mentions, contribution directions, and feature ordering from the narrative text. Compared to deterministic extraction, this approach is more robust when attribution information is conveyed implicitly rather than through explicit SHAP terminology or rank markers.

The two extraction strategies are intentionally complementary: deterministic extraction provides a fully reproducible and easily verifiable pipeline, while the LLM-based extractor provides stronger semantic interpretation capabilities.

Finally, each feature is annotated as either \textit{actionable} or \textit{non-actionable}. Using these annotations, we compute the average rank displacement between the original feature ranking and the generated narrative. Positive displacement indicates feature promotion, while negative displacement indicates demotion. This allows us to assess whether stakeholder-oriented narratives prioritize actionable information over strict adherence to the original SHAP ranking.

\nbold{Semantics.} To quantify the semantic alignment of a generated narrative with a specific target audience, we developed a reference-based evaluation approach inspired by BERTScore~\cite{zhang2019bertscore}. We first compiled a corpus of representative texts tailored to our three user profiles: clinicians, data scientists, and patients. For clinicians and data scientists, we extracted the "Results" and "Discussion" sections from scientific papers published between 2024 and 2025 on PubMed and IEEE Xplore, respectively. The search queried titles containing "diabetes" or "stroke" alongside "Machine
Learning". For the patient profile, we sourced informative lay-language medical articles from MedlinePlus\footnote{\url{https://medlineplus.gov/diabetes.html}}, the
public health information platform maintained by the U.S. National Institutes of Health (NIH). After stripping markdown artifacts and citation patterns from the collected texts, we split the narratives into individual sentences and computed their embeddings using Sentence-BERT~\cite{reimers-gurevych-2019-sentence}. This
yielded a domain-specific reference database comprising three
stakeholder-specific embedding sets for each disease.

To ensure the quality of the reference data, we refined each set by removing noisy observations. Specifically, we computed a local density score for each sentence embedding based on the mean cosine similarity to its $k=10$ nearest neighbors. Embeddings with a density Z-score outside the $[-2, 2]$ range were discarded as outliers. Finally, to prevent bias during evaluation, we randomly
undersampled the filtered sets so that all three stakeholder databases within a given disease domain contained the exact same number of embeddings.

To evaluate a generated narrative, we embed all its constituent sentences and project them into the corresponding disease's reference space. For each sentence embedding, we retrieve the $k=10$ nearest neighbors across the combined stakeholder databases. We then calculate the proportion of neighbors belonging to each audience class, averaged over all sentences in the narrative. The narrative is classified into the stakeholder category with the highest average
score. If this assigned category matches the intended target audience, the prediction is considered correct. We refer to this metric as the Semantic Accuracy and applied the evaluation exclusively to \model~and the B2 baselines, as they target multiple stakeholders.

\subsubsection{Extra-narrative Evaluation Module.} While the Intra-narrative Evaluation Module assesses the intrinsic properties of the generated texts, the Extra-narrative Evaluation Module investigates the practical utility and subjective reception of the narratives from the perspective of the user. To achieve a comprehensive and robust assessment, we structured this module around a dual evaluation strategy: an LLM-as-a-Judge approach and a Human Evaluation through surveys.

\nbold{LLM-as-a-judge.} In the LLM-as-a-judge paradigm~\cite{DBLP:conf/nips/ZhengC00WZL0LXZ23,DBLP:conf/emnlp/LiuIXWXZ23,DBLP:conf/acl/ChiangL23}, a Large Language Model is employed to assess system outputs. In our study, the judge evaluates the narratives generated by \model~and the baselines through two complementary approaches, which we refer to as Re-identification Analysis and Competitive Ranking Analysis. For these evaluations, we leveraged three distinct generative models: Claude Sonnet 4.6, GPT-5.5, and Gemini 3.1. The judge models do not share the same architectural family as the qwen or gemma backbones, structurally avoiding self-preference bias~\cite{DBLP:conf/nips/PanicksseryBF24,DBLP:conf/nips/ZhengC00WZL0LXZ23,DBLP:conf/acl/WangLCCZLCKLLS24}.




\textit{Re-identification Analysis}: To validate audience targeting, the three narratives must differ substantively in content and framing, not merely in surface markers such as headings or tone. In this task, the judge receives \model's three narratives in anonymised slots and assigns each to its intended audience. Before evaluation, narratives are sanitized to remove any system-identifying markers while preserving all prose, style, quantitative values, and SHAP terminology. The three narratives are shuffled randomly within each case, so a fixed-position guesser achieves chance baseline. Judges cannot abstain or mark ties. Across 3 judges and 208 cases, each judge evaluates 624 narrative trials.

\textit{Competitive Ranking Analysis}: For a given sample and target audience, each judge ranks all 14 narratives produced by the evaluated systems from best to worst. This task is performed independently for each of the three target audiences and across all samples in both datasets, yielding a total of $1,872$ ranking tasks and $26,208$ rank observations.

We aggregate rankings into an Elo-style score using the Bradley-Terry model~\cite{bradley1952rank,Hunter2003MMAF}, which expresses system quality as a continuous score based on pairwise win probability: a higher-scored system wins head-to-head comparisons more often. This approach is standard in LLM leaderboards~\cite{DBLP:conf/nips/ZhengC00WZL0LXZ23}. Since our pool has no canonical reference, we centre scores at the pool mean, so 0 represents the average system, and positive/negative values denote above- and below-average performance.
To determine whether differences among systems are statistically significant, we apply the Friedman omnibus test to assess whether ranking differences exist overall, and the Nemenyi post-hoc pairwise comparison~\cite{DBLP:journals/jmlr/Demsar06} to identify which specific pairs differ significantly.

Finally, to evaluate the agreement among different judges, we compute the pairwise Kendall's $\tau$~\cite{kendall} and the ordinal Krippendorff's $\alpha$~\cite{krippendorff2011computing}. The latter is a multi-rater coefficient that aggregates the consensus of all three judges into a single metric per target\footnote{Gwet's AC2 with quadratic ordinal weights~\cite{gwet2008ac1} aligns with $\alpha$ within $\pm 0.001$.}. For both metrics, higher values indicate a stronger consensus among the LLM judges.


\nbold{Human Evaluation.} To assess how human audiences perceive the \model~narratives, we conducted a preliminary evaluation through structured surveys. We randomly selected 3 instances from each dataset, resulting in a set of 6 tailored narratives per stakeholder profile (3 for Stroke, 3 for Diabetes). We recruited 15 human evaluators, divided evenly across the three audience groups (5 participants per group), to evaluate the narratives corresponding to their specific profile.  The clinician group consisted of medical specialists with at least three years of clinical experience, the data scientist group of professionals with at least three years of experience in data analysis and machine learning, and the patient group of lay participants without medical or technical backgrounds.
The assessment covered five dimensions: Understandability, Clarity, Usefulness, Trust, and Cognitive Load. The Cognitive Load dimension was adapted from the NASA Task Load Index (NASA-TLX)~\cite{hart1988development}, while the remaining items were derived from the Explanation Goodness Checklist and Explanation Satisfaction Scale~\cite{hoffman2023measures}. All dimensions were slightly adapted to fit our scenario and rated on a 5-point Likert scale (ranging from 1 = Strongly Disagree to 5 = Strongly Agree). Table~\ref{tab:human_questions} details the specific dimensions and statements presented to the patient group. Finally, to mitigate potential ordering bias, both the sequence of the medical domains (Diabetes and Stroke) and the presentation order of the narratives within each domain were fully randomized for each participant.

\begin{table}[t]
    \centering
    \small
    \caption{Evaluation dimensions and their corresponding survey statements for the patient audience. Participants rated each statement on a 5-point Likert scale.}
    \begin{tabular}{lp{5.8cm}}
    \toprule
    \textbf{Dimension} & \textbf{Statement} \\ 
    \midrule
    Understandability & I understood the factors influencing the model prediction. \\
    Clarity & The explanation was clear and easy to read. \\
    Usefulness & This explanation would help me better understand my health situation. \\
    Trust & The explanation increased my trust in the AI system. \\
    Cognitive Load & The explanation required too much effort to understand. \\
    \bottomrule
    \label{tab:human_questions}
    \end{tabular}
\end{table}

\section{Results and Discussion}
\subsection{Intra-narrative analysis}
\subsubsection{Form Analysis}
\begin{table}[t]
\centering
\setlength{\tabcolsep}{3pt} 
\small
\caption{Form-analysis metrics averaged across both datasets. Lower/higher values are preferred as indicated by the arrows.}
\label{tab:form_analysis}
\setlength{\tabcolsep}{4pt}
\begin{tabular}{llcccc}
\toprule
\textbf{Model} & \textbf{Audience} & \textbf{\#Words} & \textbf{FKGL} $\downarrow$ & \textbf{MATTR} $\uparrow$ & \textbf{Rep-3} $\downarrow$ \\
\midrule

B1-gemma & Generic & 166.56 & 10.76 & 0.628 & 0.206 \\
B1-qwen & Generic & 214.68 & 14.84 & 0.624 & 0.238 \\

B2-gemma & Clinician & 157.45 & 11.42 & 0.704 & 0.117 \\
B2-gemma & Data Scientist & 177.84 & 8.54 & 0.640 & 0.189 \\
B2-gemma & Patient & 153.93 & 10.35 & 0.782 & 0.081 \\

B2-qwen & Clinician & 191.01 & 14.88 & 0.716 & 0.135 \\
B2-qwen & Data Scientist & 269.44 & 14.07 & 0.668 & 0.163 \\
B2-qwen & Patient & 162.13 & 12.79 & 0.812 & 0.122 \\

B3 & Generic & 215.63 & 13.66 & 0.601 & 0.362 \\
B4 & Generic & 172.58 & 10.98 & 0.574 & 0.296 \\

Explingo & Generic & 38.38 & 15.37 & 0.792 & 0.001 \\

\model & Clinician & 150.77 & 17.06 & 0.787 & 0.129 \\
\model & Data Scientist & 232.23 & 15.69 & 0.752 & 0.042 \\
\model & Patient & 161.63 & 12.09 & 0.806 & 0.064 \\

\bottomrule
\end{tabular}
\end{table}
Table~\ref{tab:form_analysis} reports the stylistic and readability-oriented metrics computed across the generated narratives. Overall, the analysis reveals substantial differences in verbosity, readability, lexical diversity, and repetition patterns across both systems and stakeholder configurations.

The results show that audience adaptation affects both the content and style of the generated explanations. Patient-oriented narratives generally exhibit lower linguistic complexity than clinician- and data scientist-oriented ones. For example, \model:P achieves a lower FKGL score ($12.09$) than \model:DS ($15.69$), while data scientist narratives tend to be longer and more detailed.

Narrative repetition also varies substantially across systems. B3 and B4 exhibit the highest repeated trigram ratios ($0.362$ and $0.296$, respectively), together with comparatively low MATTR scores, suggesting stronger reliance on template-based generation patterns. In contrast, \model~maintains consistently low repetition across all stakeholders, particularly for the data scientist target ($0.042$), despite generating longer narratives. This suggests that iterative refinement may help reduce boilerplate generation.

Lexical diversity further supports this interpretation. Our approach achieves consistently high MATTR scores across all targets, particularly for patient-oriented narratives ($0.806$), suggesting a more varied language use. Explingo also achieves high MATTR values with almost zero repetition; however, this behavior is likely influenced by its extremely short outputs ($\sim40$ words on average) rather than by a richer discourse structure. Indeed, despite its brevity, Explingo exhibits one of the highest FKGL scores in the benchmark.

The comparison between B2 and \model~ also provides insights into the role of multi-stage refinement. B2 already demonstrates that simple role prompting is sufficient to induce measurable stylistic adaptation across stakeholders. However, \model~ generally achieves higher lexical diversity together with lower repetition, suggesting that iterative refinement may help generate more natural and less templatic narratives.

Overall, \model~ combines stakeholder adaptation, lexical richness, and low repetition while maintaining substantially longer narratives than compressed approaches like Explingo.

\subsubsection{Content Analysis}
\begin{table}[t]
\centering
\setlength{\tabcolsep}{4pt} 
\small
\caption{Content-analysis metrics averaged across both datasets. Scores are reported as Deterministic (D) / LLM (L).}
\label{tab:faithfulness}
\begin{tabular}{llccc}
\toprule
\textbf{Model} & \textbf{Audience} & \textbf{VA (D/L)} & \textbf{SA (D/L)} & \textbf{RA (D/L)} \\
\midrule


B1-gemma
& Generic 
& 0.995 / 0.995 
& 0.972 / 0.995 
& 0.996 / 0.997 \\

B1-qwen
& Generic 
& 1.000 / 1.000 
& 0.972 / 1.000 
& 0.979 / 1.000 \\

B2-gemma 
& Clinician 
& 1.000 / 1.000 
& 0.917 / 1.000 
& 0.943 / 0.971 \\

B2-gemma 
& Data Scientist 
& 1.000 / 1.000 
& 0.985 / 1.000 
& 0.985 / 0.988 \\

B2-gemma 
& Patient 
& 0.972 / 0.972 
& 0.563 / 0.991 
& 0.888 / 0.891 \\

B2-qwen 
& Clinician 
& 1.000 / 1.000 
& 0.947 / 1.000 
& 0.963 / 1.000 \\

B2-qwen 
& Data Scientist 
& 1.000 / 1.000 
& 0.989 / 1.000 
& 0.991 / 1.000 \\

B2-qwen 
& Patient 
& 0.650 / 0.782 
& 0.337 / 0.992 
& 0.926 / 0.961 \\

B3 
& Generic 
& 1.000 / 1.000 
& 0.735 / 1.000 
& 1.000 / 1.000 \\

B4 
& Generic 
& 1.000 / 0.999 
& 0.802 / 1.000 
& 0.993 / 0.999 \\

Explingo 
& Generic 
& 1.000 / 0.994 
& 0.075 / 0.999 
& 0.855 / 0.871 \\

\model 
& Clinician 
& 1.000 / 0.998 
& 0.979 / 1.000 
& 0.897 / 0.989 \\

\model 
& Data Scientist 
& 0.995 / 0.994 
& 0.892 / 1.000 
& 0.953 / 0.988 \\

\model 
& Patient 
& 0.000 / 0.000 
& 0.376 / 0.698 
& 0.773 / 0.789 \\

\bottomrule
\end{tabular}
\end{table}

Table~\ref{tab:faithfulness} reports the Value Agreement (VA), Sign Agreement (SA), and Rank Agreement (RA) scores computed using both deterministic and LLM-based extraction approaches.

Regarding Value Agreement, most systems achieve nearly perfect scores under both extraction strategies, indicating that feature values are generally preserved correctly whenever they are explicitly mentioned in the narrative. The main exception concerns patient-oriented explanations, where raw feature values are intentionally suppressed. Interestingly, the B2 patient configuration still achieves non-negligible VA scores ($0.650/0.782$), suggesting that role prompting alone only partially suppresses explicit feature values. In contrast, \model:P consistently suppresses explicit value mentions, indicating that the refinement stages enforce stakeholder-oriented communication constraints more effectively.

For both Sign Agreement and Rank Agreement, the LLM-based extraction strategy generally achieves higher scores than deterministic extraction, particularly for Explingo and patient-oriented narratives. In particular, SA scores obtained with LLM extraction are close to or equal to $1.0$ in most configurations, indicating that the direction of feature contributions is usually preserved even when attribution information is conveyed implicitly. Deterministic extraction, in contrast, produces lower scores in settings where explanatory information is expressed through free-form language rather than explicit SHAP terminology or numeric values.

Rank Agreement also reveals the impact of narrative rewriting on explanation structure. Explingo exhibits the lowest RA values ($0.855/0.871$), suggesting that aggressive rewriting substantially alters the ordering of explanatory evidence. \model~ instead maintains stronger alignment with the original SHAP ranking for clinician and data scientist narratives, while patient-oriented explanations intentionally promote \emph{actionable} features, i.e., features that the patient can potentially modify or intervene on. For example, body mass index is considered actionable, whereas age is not. This design prioritizes intervention-oriented communication over strict adherence to the original SHAP ordering.

\begin{table}[t]
\centering
\small
\caption{Average feature rank displacement from the original ranking. Positive values indicate promotion in the generated narrative, negative values demotion.}
\label{tab:actionability_displacement}
\begin{tabular}{llcc}
\toprule
\textbf{Model} & \textbf{Audience} & \textbf{Actionable} & \textbf{Non-actionable} \\
\midrule
\model & Clinician       & 0.011  & -0.006 \\
\model & Data Scientist  & 0.004  & 0.013 \\
\rowcolor{gray!20}
\model & Patient         & 0.622  & -0.305 \\
\bottomrule
\end{tabular}
\end{table}

To validate this hypothesis, Table~\ref{tab:actionability_displacement} reports the average displacement of actionable and non-actionable features with respect to their original ranking. A clear divergence emerges for patient-oriented narratives: actionable features are promoted on average by $+0.622$ positions, while non-actionable features are demoted by $-0.305$. This effect is nearly absent for clinician and data scientist narratives, confirming that the patient-oriented strategy systematically prioritizes actionable information.

\subsubsection{Semantic Analysis} The results of the semantic evaluation are detailed in Table~\ref{tab:semantic}. Across both datasets, \model~significantly outperforms the B2 implementations (\texttt{gemma} and \texttt{qwen}) for the Patient and Clinician audiences. Notably, the baseline models struggle severely to adapt their language for the Patient, yielding semantic accuracies consistently below 0.30 on the diabetes dataset and below 0.10 on the stroke dataset.

B2 achieves perfect accuracy (1.00) on the stroke dataset for the Data Scientist, slightly surpassing \model~(0.92). However, the full approach maintains a much more robust and balanced performance across all stakeholder profiles. Consequently, the overall accuracy of \model~is more than double that of the baselines on both datasets (0.89 vs. 0.44/0.42 for stroke, and 0.83 vs. 0.41/0.35 for diabetes).

These findings highlight a critical limitation of standard audience-aware generation: simply embedding a role description into an LLM prompt (as done in B2) is insufficient to produce narratives with accurately tailored semantics. This gap is especially pronounced for the patient stakeholder, who represents an audience external to the professional medical and technical domains. Conversely, the introduction of a Reviewer Agent in our architecture appears effective, enabling both the generation and the targeted refinement of narratives which semantically match their intended stakeholders.

\begin{table}\label{semantic}
\centering
\caption{Semantic accuracy across datasets and stakeholder targets. The best results for each dataset are in bold.}
\small
\setlength{\tabcolsep}{3.5pt}
\begin{tabular}{llcccc}
\toprule
\textbf{Dataset} & \textbf{Approach} & \textbf{Patient} & \textbf{Clinician} &\textbf{Data Scientist} & \textbf{Overall} \\
\midrule
\multirow{3}{*}{Stroke} & B2-gemma & 0.04 & 0.29 & \textbf{1.00} & 0.44 \\
& B2-qwen & 0.01 & 0.24 & \textbf{1.00} & 0.42 \\
& \model & \textbf{0.79} & \textbf{0.95} & 0.92 & \textbf{0.89} \\    
\midrule
\multirow{3}{*}{Diabetes} & B2-gemma & 0.27 & 0.46 & 0.50 & 0.41 \\
& B2-qwen & 0.21 & 0.57 & 0.27 & 0.35 \\
& \model & \textbf{0.99} & \textbf{0.94} & \textbf{0.56} & \textbf{0.83} \\
\bottomrule
\label{tab:semantic}
\end{tabular}
\end{table}

\subsection{Extra-Narrative Analysis} 
\subsubsection{LLM-as-a-judge: Re-identification Analysis}\label{sec:eval}\noindent
All three judges consistently assigned every generated narrative to its intended audience on both datasets, reaching $100\%$ per-narrative accuracy and $100\%$ triplet full-match. These scores are well above the corresponding random baselines of $33\%$ and $16.7\%$. This suggests that \model~does not merely alter surface markers such as headings or tone, but produces narratives whose content and framing are consistently recognizable as audience-specific.

While these scores show that our approach tailors narratives to their intended audiences, they may also reflect the advanced reasoning capabilities of the state-of-the-art LLMs acting as judges, combined with the relative simplicity of the discriminative task. To provide a more rigorous and granular evaluation, we complement this analysis with the Competitive Ranking Analysis, directly comparing the quality of \model~against the baselines in a competitive scenario.

\subsubsection{LLM-as-a-judge: Competitive Ranking Analysis.}\label{sec:eval:modB}

\begin{figure*}[t]
\centering
\includegraphics[width=\textwidth]{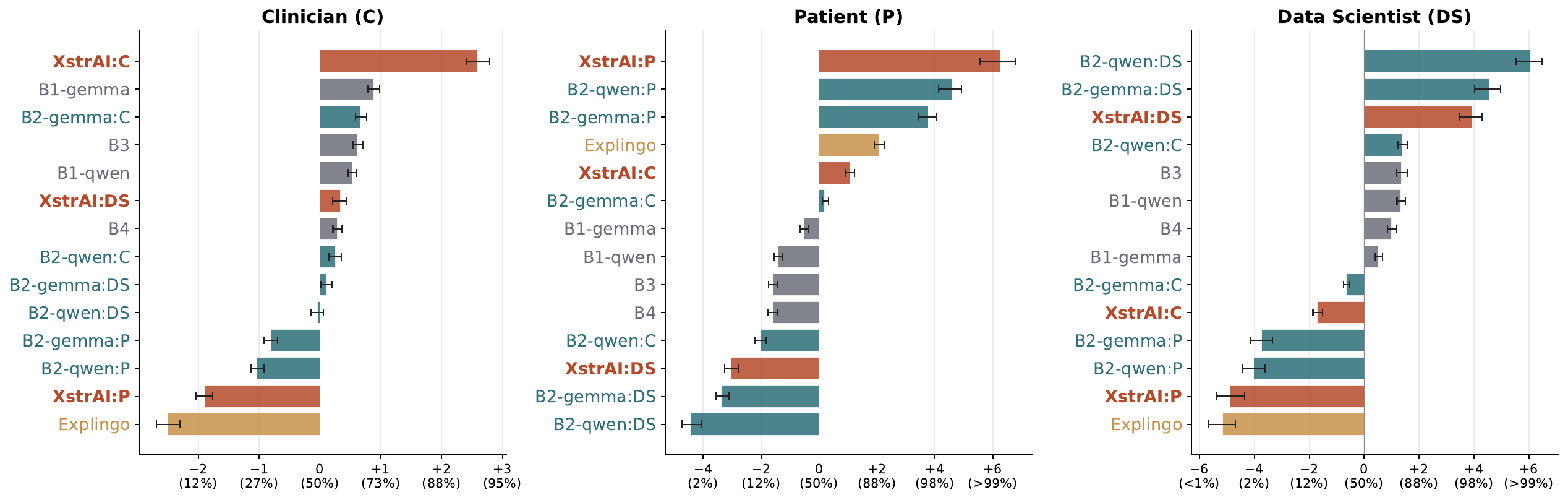}
\caption{Competitive ranking analysis: Elo log-strength (zero-centered; 0 = average system) across three audiences, pooled over 3 judges and 2 datasets. Panels are sorted by Elo score, and percentages indicate pairwise win probability against the average system (e.g., Elo $+1 \approx 73\%$). Error bars represent $95\%$ bootstrap confidence intervals.}
\label{fig:bt}
\end{figure*}

This analysis evaluates whether systems are preferred by LLM judges when narratives are ranked for a specific target audience. Figure~\ref{fig:bt} reports Bradley-Terry Elo scores with confidence intervals per audience; Table~\ref{tab:kendall} reports inter-judge agreement.

The Elo scores are associated with narrow error bars, indicating stable ranking differences. The Friedman test rejects the null hypothesis that all systems are equivalent ($p\!\ll\!10^{-10}$ on every panel), and the Nemenyi pairwise comparison sets a critical difference of $0.79$ rank positions, a threshold exceeded by most observed gaps.

\model~ ranks first on two of its three matched audiences: \textit{Clinician} (Elo $+2.60$, $\sim\!85\%$ wins over {B1-gemma}) and \textit{Patient} (Elo $+6.27$, $\sim\!84\%$ wins over {B2-qwen:P}). On \textit{Data Scientist}, it ranks third behind {B2-qwen:DS} (Elo $+6.07$) and {B2-gemma:DS} (Elo $+4.56$).

A consistent specialization effect is visible across all three panels. Audience-conditioned systems dominate their matched target but score poorly elsewhere: {\model:P}, for instance, moves from $+6.27$ on \textit{Patient} to $-4.88$ on \textit{Data Scientist}. Generic baselines (B1, B3, B4) remain near the pool average regardless of target, confirming that audience-specific adaptation is a necessary component of narrative quality. Explingo ranks among the weakest overall, performing comparatively better on \textit{Patient} alone, where its short and direct style partially satisfies accessibility requirements. Its consistently low scores on \textit{Clinician} and \textit{Data Scientist} show that concision alone cannot substitute for controlled, attribution-aware generation.

\begin{table}[t]
\centering
\caption{Inter-judge agreement on the Competitive Ranking Analysis. Pairwise mean Kendall~$\tau$ (averaged over the two datasets) and ordinal Krippendorff's $\alpha$ per dataset (multi-rater across all 3 judges). Higher $=$ more agreement; the shaded row falls below the 0.80 reliability cut-off.}
\label{tab:kendall}
\small
\setlength{\tabcolsep}{6pt}
\begin{tabular}{lccccc}
\toprule
& \multicolumn{3}{c}{\textbf{Mean Kendall} $\boldsymbol\tau$} & \multicolumn{2}{c}{\textbf{Krippendorff} $\boldsymbol{\alpha}_{\text{ord}}$} \\
\cmidrule(lr){2-4}\cmidrule(l){5-6}
\textbf{Audience} & \shortstack{Claude\\GPT} & \shortstack{Claude\\Gemini} & \shortstack{GPT\\Gemini} & \textit{diabetes} & \textit{stroke} \\
\midrule
Patient        & 0.80 & 0.70 & 0.71 & 0.87 & 0.82 \\
Data Scientist & 0.85 & 0.72 & 0.73 & 0.90 & 0.86 \\
\rowcolor{gray!20}
Clinician & 0.23 & 0.31 & 0.33 & 0.34 & 0.45 \\
\bottomrule
\end{tabular}
\end{table}

\nbold{Audience-level interpretation.} The three audiences reveal distinct patterns in how much pipeline complexity the task demands.

On \textit{Clinician}, inter-judge agreement is the lowest of the three targets (Table~\ref{tab:kendall}), reflecting the legitimate diversity of acceptable clinical styles rather than evaluator noise. Yet it is precisely on this contested target that {\model:C} achieves its widest margin in the benchmark: a $+1.71$ Elo lead over the runner-up {B1-gemma}, corresponding to a $\sim\!85\%$ pairwise win probability and the largest first-to-second gap observed across all panels. The result suggests that multi-round refinement is most valuable where no single formulation dominates, allowing the pipeline to converge on outputs that single-prompt baselines, committed to one style, cannot match.

On \textit{Patient}, {\model:P} leads, followed by the baselines {B2-qwen:P} and {B2-gemma:P}. Explingo performs better here than on any other target, confirming that simplified, accessible language partially meets patient requirements. Even so, audience-conditioned systems rank above it, showing that simplicity alone is insufficient: a personalized and carefully framed explanation remains preferable to a merely concise one. The stronger inter-judge consensus on this target (ordinal $\alpha \geq 0.82$) confirms the stability of this ordering.

On \textit{Data Scientist}, the dominant requirement is raw SHAP attribution density and quantitative completeness. A data scientist familiar with feature-attribution methods expects explicit values, rankings, and directions, properties that a single dense prompt naturally maximises. {B2-qwen:DS} and {B2-gemma:DS} lead accordingly, with {\model:DS} close behind. All top-ranked systems are audience-conditioned, confirming that specialization remains necessary even when the narrative structure is simpler. The gap between audience-conditioned and generic systems is as large here as on any other target; what changes is the form that 
specialization should take.

\begin{table}[t!]
    \centering
    \small
        \caption{Average human evaluation scores on a 5-point Likert scale (1 = Strongly Disagree, 5 = Strongly Agree). Higher scores are better, except for Cognitive Load.}
    \begin{tabular}{lccc}
    \toprule
    \textbf{Dimension} &  \textbf{Patient} & \textbf{Data Scientist} & \textbf{Clinician} \\
    \midrule
    Understandability & 4.2 & 3.9 & 4.7 \\
    Clarity & 3.8 & 3.8 & 4.3 \\
    Usefulness & 3.5 & 3.6 & 4.3 \\
    Trust & 3.0 & 3.3 & 4.1 \\
    Cognitive Load & 2.2 & 2.4 & 2.4 \\
    \bottomrule
    \end{tabular}
    \label{tab:human_eval_results}
\end{table}

\subsubsection{Human Evaluation.}\label{sec:eval_human} 

Table~\ref{tab:human_eval_results} reports the average human evaluation scores across the five assessed dimensions for each target audience. Clinicians reported the highest scores among all stakeholders for Understandability (4.7), Clarity (4.3), Usefulness (4.3), and Trust (4.1), indicating that the generated narratives were exceptionally well-received and deemed highly supportive by medical professionals.

Interestingly, the Data Scientist group assigned more modest ratings across these dimensions, with average scores ranging between 3 (Neutral) and 4 (Agree). As detailed in Table~\ref{tab:form_analysis}, this group received the longest narratives, with an average of 232 words compared to 150 for clinicians. Although they processed longer texts, their reported Cognitive Load was identical to that of the medical audience (2.4). This suggests that the increased length did not translate into a significantly higher reading effort for technical users.

Finally, Patients found their tailored narratives to be highly understandable (4.2) and clear (3.8), while experiencing the lowest Cognitive Load among all groups (2.2). However, there is room for improvement regarding Trust (3.0) and Usefulness (3.5), which received neutral-to-positive, yet limited, evaluations.

These findings suggest that \model~generates audience-specific narratives that are clear and require low cognitive effort. Given the limited number of participants, these findings should be interpreted as a preliminary but promising assessment of stakeholder-specific narrative perception rather than as a definitive usability validation.

\section{Conclusion}
This paper introduced \model, a multi-agent framework for generating stakeholder-specific narratives from fixed local feature-attribution evidence. By separating explanatory evidence from its communication, \model~employs an immutable \texttt{ExplanationCard} and a Framer--Narrator--Reviewer pipeline to control how the same prediction explanation is selected, organized, verbalized, and assessed for different audiences.

Our evaluations show that \model~produces clearly differentiated narratives while showing strong empirical alignment with the underlying attribution evidence. Compared to generic and single-prompt baselines, the framework achieves lower repetition, higher lexical diversity, stronger audience fit, and competitive or superior performance in deterministic and LLM-based assessments. The findings also indicate that stakeholder adaptation may legitimately reshape explanation structure; for instance, in patient narratives, actionable features are prioritized over strict SHAP-rank ordering.

Future work will examine prompt-level ablations and the robustness of the framework under controlled noise in the input explanations. We also plan to assess whether the same communication layer design generalizes beyond SHAP to other XAI techniques and beyond tabular clinical prediction to other application contexts. Finally, larger and more targeted human evaluations will be needed to assess stakeholder-specific usefulness, trust, cognitive load, and communicative adequacy more deeply.

\bibliographystyle{ACM-Reference-Format}
\bibliography{bibliography}

\end{document}